\documentclass[aps,prl,preprint,superscriptaddress]{revtex4-1}
\usepackage{graphicx}% Include figure files
\usepackage{dcolumn}% Align table columns on decimal point
\usepackage{bm}% bold math
\usepackage{amssymb,amsmath}
\usepackage[british]{babel}
\usepackage{natbib}

\setcitestyle{super}

\begin{document}

% Use the \preprint command to place your local institutional report
% number in the upper righthand corner of the title page in preprint mode.
% Multiple \preprint commands are allowed.
% Use the 'preprintnumbers' class option to override journal defaults
% to display numbers if necessary
%\preprint{}

%Title of paper
\title{Vectorial optical field reconstruction by attosecond spectral interferometry.}

% repeat the \author .. \affiliation  etc. as needed
% \email, \thanks, \homepage, \altaffiliation all apply to the current
% author. Explanatory text should go in the []'s, actual e-mail
% address or url should go in the {}'s for \email and \homepage.
% Please use the appropriate macro foreach each type of information

% \affiliation command applies to all authors since the last
% \affiliation command. The \affiliation command should follow the
% other information
% \affiliation can be followed by \email, \homepage, \thanks as well.
\author{P.~Carpeggiani$^{1,2\dag}$, M.~Reduzzi$^{1,2\dag}$, A.~Comby$^{1\dag}$, H.~Ahmadi$^{1,3}$, S.~K\"{u}hn$^{4}$, F.~Calegari$^{2,5,6}$, M.~Nisoli$^{1,2}$, F.~Frassetto$^{7}$, L.~Poletto$^{7}$, D.~Hoff$^{8}$, J.~Ullrich$^{9}$, C.~D.~Schr\"{o}ter$^{10}$, R.~Moshammer$^{10}$, G.~G.~Paulus$^{8}$, G.~Sansone$^{1,2,4,11}$.}
\affiliation{(1) Dipartimento di Fisica, Politecnico Piazza Leonardo da Vinci 32, 20133 Milano Italy\\
(2) IFN-CNR, Piazza Leonardo da Vinci 32, 20133 Milano Italy\\
(3) Department of Physical Chemistry, School of Chemistry, College of Science, University of Tehran, Tehran, Iran\\
(4) ELI-ALPS, ELI-Hu Kft., Dugonics ter 13, H-6720 Szeged, Hungary \\
(5) Deutsches Elektronen-Synchrotron, Notkestrasse 85, Hamburg 22607, Germany\\
(6) Physics Department, University of Hamburg, Luruper Chaussee 149, 22761 Hamburg, Germany\\
(7) Institute of Photonics and Nanotechnologies, CNR via Trasea 7, 35131 Padova Italy\\
(8) Institut f\"{u}r Optik und Quantenelektronik, Friedrich-Schiller-Universit\"{a}t Jena, Max-Wien-Platz 1, 07743 Jena, Germany\\
(9) Physikalisch-Technische Bundesanstalt, Bundesallee 100, 38116 Braunschweig, Germany\\
(10) Max-Planck-Institut f\"{u}r Kernphysik, Saupfercheckweg 1, 69117 Heidelberg, Germany\\
(11) Physikalisches Institut, Albert-Ludwigs-Universit\"{a}t Freiburg, 79106 Freiburg, Germany\\
$\dag$ These authors contributed equally to this work}

%Collaboration name if desired (requires use of superscriptaddress
%option in \documentclass). \noaffiliation is required (may also be
%used with the \author command).
%\collaboration can be followed by \email, \homepage, \thanks as well.
%\collaboration{}
%\noaffiliation

\date{\today}

\begin{abstract}
%~\cite{NATPHOT-Kim-2013,SCIENCE-Wirth-2011}.
An electrical pulse $\mathbf{E}(t)$ is completely defined by its time-dependent amplitude and polarisation direction. For optical pulses the manipulation and characterisation of the light polarisation state is fundamental due to its relevance in several scientific and technological fields.
In this work we demonstrate the complete temporal reconstruction of the electric field of few-cycle pulses with a complex time-dependent polarisation. Our experimental approach is based on extreme ultraviolet interferometry with isolated attosecond pulses and on the demonstration that the motion of an attosecond electron wave packet is sensitive to perturbing fields only along the direction of its motion. By exploiting the sensitivity of interferometric techniques and by controlling the emission and acceleration direction of the wave packet, pulses with energies as low as few hundreds of nanojoules can be reconstructed. Our approach opens the possibility to completely characterise the electric field of the pulses typically used in visible pump-probe spectroscopy.
\end{abstract}

% insert suggested PACS numbers in braces on next line
%\pacs{}
% insert suggested keywords - APS authors don't need to do this
%\keywords{}

%\maketitle must follow title, authors, abstract, \pacs, and \keywords
\maketitle

% body of paper here - Use proper section commands
% References should be done using the \cite, \ref, and \label commands

The complete characterisation of optical pulses requires the possibility to sample in time oscillating electric fields on the sub-femtosecond timescale. In the last decade, the development of attosecond science and technology has made available the technological tools towards this goal$^{\cite{RMP-Krausz-2009, PQE-Nisoli-2009, SCIENCE-Kapteyn-2007}}$. In general, the electric field $\mathbf{E}(t)$ is a vector quantity and the measurement of the (time-dependent) polarisation is required for its complete reconstruction. The polarisation state of the radiation is a key parameter in the light-matter interaction, as it determines selection rules in photoexcitation and photoionisation of atoms and molecules, and spin properties of photoelectrons emitted from surfaces. Moreover, polarisation shaping and synthesisation of complex time-dependent polarisations are widely adopted schemes to control laser-matter interaction at high-intensities~\cite{OL-Corkum-1994, NATPHYS-Eckle-2008}, at the nanoscale level~\cite{NAT-Aeschlimann-2011}, and for the optimal control of quantum systems~\cite{PRL-Brixner-2004}.\\
An electric field with a time-dependent polarisation can be retrieved by measuring its components along two (perpendicular) directions. For ultrashort pulses approaching the single-cycle regime~\cite{SCIENCE-Hassan-2016}, the rotation of the polarisation axis through wave plates can introduce distortions due to the large spectral bandwidth and higher-order phase dispersion terms. Therefore, the characterisation along these two directions should avoid the use of such additional dispersive elements.\\
A technique capable to fully characterise weak probe pulses with energy of a few tens or hundreds of nanojoules would be also desirable. In this way the full reconstruction of electric fields would not be limited to intense pulses~\cite{NAT-Schultze-2016}, but it could be extended to the probe pulses typically used in visible and infrared pump-probe spectroscopy.\\
Two complementary approaches have been demonstrated so far for the complete characterisation of optical waveforms: the attosecond streak camera~\cite{PRL-Itatani-2002, PRL-Kitzler-2002} and the petahertz optical oscilloscope~\cite{NATPHOT-Kim-2013}. Electric fields with linear~\cite{Science-Goulielmakis-2004} and time-dependent polarisation~\cite{OE-Boge-2014} were retrieved using the first approach, which, however, requires the use of intense fields. On the other hand, the second technique introduces a systematic distortion of the reconstructed waveform due to the non-collinear geometry~\cite{NATPHOT-Kim-2013}.
\subsection{XUV spectral interferometry with isolated attosecond pulses}
In our experimental approach we take advantage of extreme ultraviolet (XUV) spectral interferometry driven by isolated attosecond pulses to demonstrate the full reconstruction of unknown fields characterised by a complex polarisation pattern.
XUV spectral interferometry of high-order harmonic sources was already demonstrated with trains of attosecond pulses for the investigation of plasma~\cite{PRL-Salieres-1999} and molecular~\cite{PRA-Camper-2014} dynamics. In these approaches one XUV pulse acts as a reference, while the characteristics (phase or amplitude) of the second one are modified by the ongoing dynamics.\\
\begin{figure}[p]
\centering\includegraphics[width=16cm]{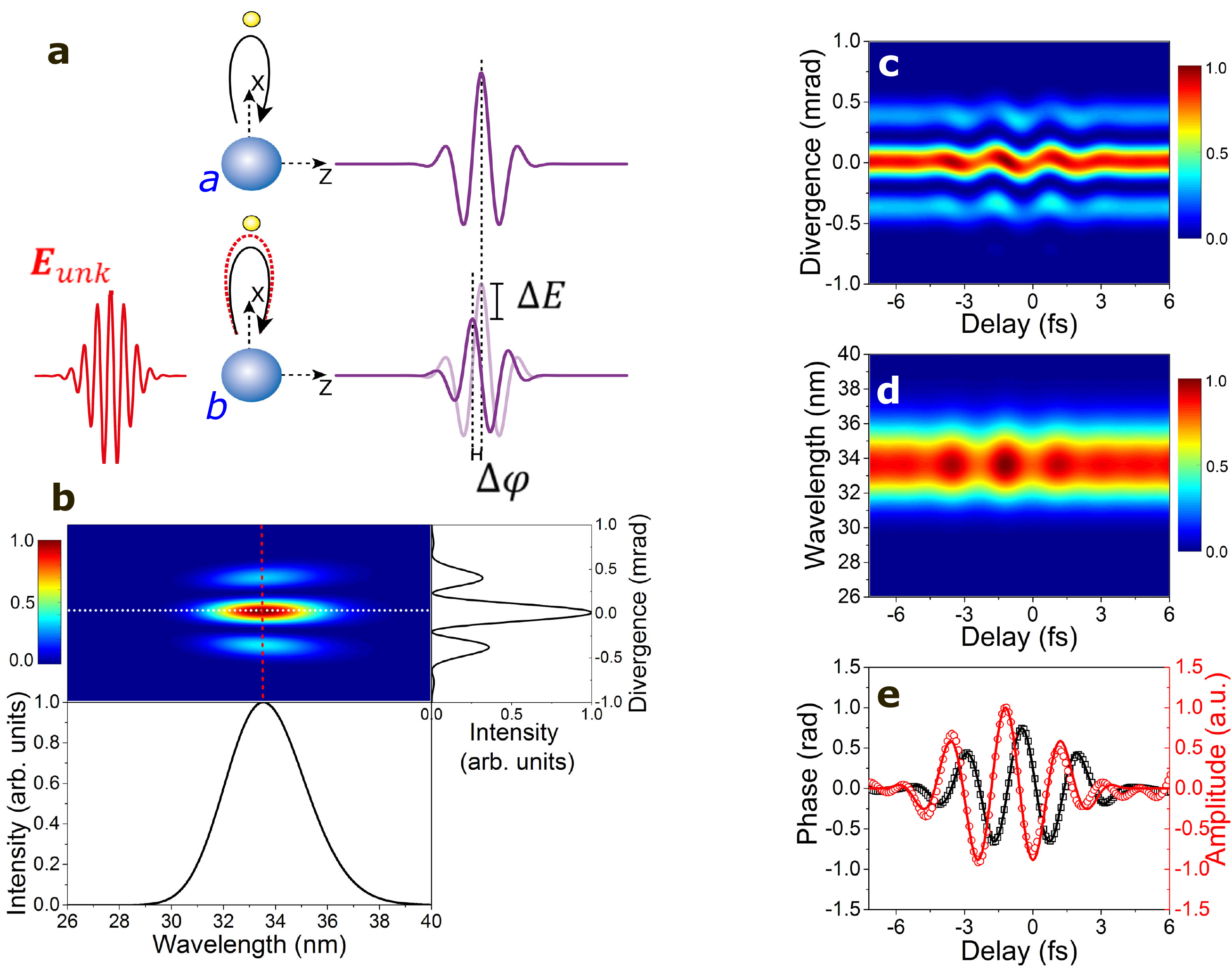}
\caption{\textbf{XUV spectral interferometry with isolated attosecond pulses.}\\
\textbf{a}, Two electronic wave packets, upon recollision, lead to the generation of two coherent isolated attosecond pulses, which interfere in the far-field. A weak probe pulse $\mathbf{E}_{unk}$ perturbs the amplitude ($\Delta E$) and phase ($\Delta\varphi$) of one of the two attosecond pulses. \textbf{b}, Simulated XUV spectral interference pattern on the output plane of an XUV spectrometer. The right and lower panel show the projections along the vertical (red dashed line) and horizontal (white dashed line) directions, corresponding to the divergence and the spectrum of the XUV interference pattern, respectively. The parameters of the experiment have been used in the simulations (see Supplementary Information). \textbf{c},\textbf{d}, Dependence of the divergence (\textbf{c}) and of the spectrum (\textbf{d}) of the simulated XUV interference pattern as a function of the delay $\tau$. \textbf{e}, Electric fields retrieved from the amplitude (red circles, arbitrary units) and the phase (black squares) using a Fourier analysis of the oscillations of Fig.~\ref{Fig1}c (see Supplementary Information). The time difference between the two reconstructed fields is about a quarter of optical cycle. The field used in the simulation is shown by solid line (red and black). The input fields have been shifted and normalised to match the reconstructed ones. Parameters used in the simulations: FWHM=5~fs, $\lambda$=744~nm, $I=3.5\times 10^{14}~\mathrm{W/cm^2}$, CEP=0 for the driving field; FWHM=5~fs, $\lambda$=744~nm, $I=3.5\times 10^{10}~\mathrm{W/cm^2}$, CEP=0 for the unknown field.}
\label{Fig1}
\end{figure}
The application of XUV interferometry for the reconstruction of optical pulses is presented in Fig.~\ref{Fig1}a: two coherent intense driving fields are focused in two closely-spaced focal spots, where two attosecond electronic wave packets are released by tunnel ionisation in the continuum and accelerated (a,b). When the electric field drives back the wave packets, two closely-spaced coherent isolated attosecond pulses are generated upon recombination of the electrons with the ground state of the parent ions. The polarisation direction of the isolated attosecond pulses is linked to the direction of motion of the electronic wave packets ($x$ direction in Fig.~\ref{Fig1}a), which is determined by the direction of polarisation of the driving field. The two XUV pulses interfere in the far-field along the vertical direction, while the spectra are dispersed by a suitable optical setup in the perpendicular direction~(Fig.~\ref{Fig1}b). The unknown, perturbing optical pulse $\mathbf{E}_{unk}$ is overlapped in one of the two focal spots. The weak field perturbs the generation of XUV radiation leading to a modulation in the amplitude ($\Delta E$) and phase ($\Delta\varphi$) of one of the two isolated attosecond pulses. This modulation is imprinted in the strength (amplitude) and position (phase) of the interference fringes (see Movie~1 in the Supplementary Information), and it is determined by the instantaneous unknown field during the motion of the electronic wave packet~\cite{NATPHOT-Kim-2013}.
Indeed $\Delta E$ and $\Delta\varphi$ are related to the imaginary and real part of the phase accumulated by the electron wave packet in the continuum ($S$)~\cite{PRA-Lewenstein-1994}, which is determined by the ionisation ($t_{ion}$) and recombination ($t_{rec}$) times, and by the total field experienced by the electron wave packet in the continuum. We have verified that these three quantities depend linearly on the amplitude of the unknown field (in a suitable intensity range, see discussion below).\\
By changing the arrival time $\tau$ of the pulse, the fringe position and the total signal of the XUV spectrum oscillate as shown in Figs.~\ref{Fig1}c and~\ref{Fig1}d, respectively. The electric field of the perturbing field can be reconstructed from either the amplitude or the phase of the oscillation of the interference pattern, as shown in Fig.~\ref{Fig1}e, which reports the input field (solid lines) and the electric field reconstructed from the amplitude and phase of the oscillations. The input field was shifted in time and normalised to match the reconstructed fields.
The amplitude modulation of the interference pattern can be attributed to the variation of the ionisation probability, due to the presence of the unknown field. Therefore, the field reconstructed from this modulation corresponds to the unknown field at the ionisation instant. On the other hand, the phase modulation of the interference pattern is due to the variation of the phase accumulated by the electron wave packet between the ionisation and recombination instants. The reconstructed field corresponds to an average value of the unknown field between these two instants. For this reason, the reconstructions from the amplitude and phase modulations present a temporal offset, as shown in Fig.~\ref{Fig1}e.\\
We have verified that the effect on the phase of the interference pattern scales linearly with the electric field for peak intensities between $I=5\times10^8~\mathrm{ W/cm^2}$ and $I=9\times10^{12}~\mathrm{W/cm^2}$. The effect on the amplitude of the modulation is linear in a smaller intensity range (up to about $I=5\times10^{11}~\mathrm{ W/cm^2}$). This difference can be attributed to the exponential dependence of the tunnelling ionisation rate on the electric field, which limits the intensity range for the linear dependence on the unknown field. For this reason, in the following we will focus on the electric fields reconstructed from the phase of the interference pattern.\\
We have also verified that pulses with a carrier-wavelength as short as $200~\mathrm{nm}$ can be reproduced with an error between 5\% and 15\% (evaluated on the single reconstructed electric field value) depending on the duration of the unknown pulse.\\

\subsection{Attosecond electron dynamics as time-gated directional field detector}
\begin{figure}[htb]
\centering\includegraphics[width=10cm]{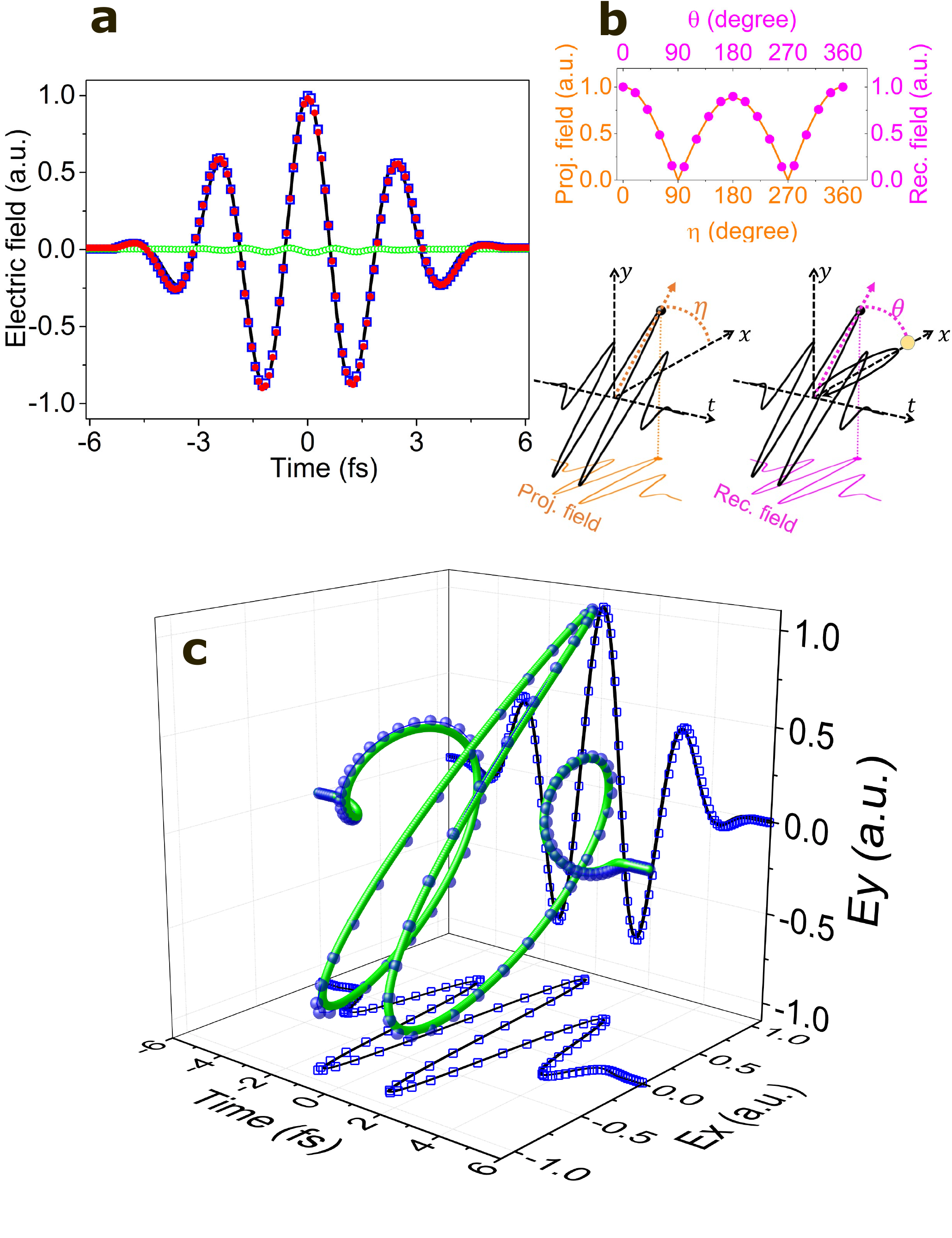}
\caption{\textbf{Single electron recollision as time-gated directional field detector.}
\textbf{a}, Simulated (black line) and reconstructed electric field in the case of linear polarisation parallel (blue open squares) and perpendicular (green open circles) to the motion of the electronic wave packet, and in the case of circular polarisation (red full circles). Parameters of the simulation: FWHM=4~fs, $\lambda$=600~nm, carrier envelope phase CEP=0. \textbf{b}, Maximum amplitude of the projection of the electric field of a linearly polarised pulse along the reference $x$ direction, as a function of the relative angle $\eta$ between this direction and the polarisation axis of the field (orange line). Maximum amplitude of the reconstructed field, as a function of the relative angle $\theta$ between the motion of the electronic wave packet and the polarisation axis of the field (magenta points). \textbf{c}, Simulated (green solid line) and retrieved electric fields (blue spheres) for a pulse with a time-dependent polarisation state. The projections along the $x$ and $y$ directions are shown for the input (black lines; $x$ direction FWHM=4~fs, $\lambda$=600~nm, $I=3.5\times 10^{10}~\mathrm{W/cm^2}$, CEP=0; $y$ direction FWHM=4~fs, $\lambda$=744~nm, $I=3.5\times 10^{10}~\mathrm{W/cm^2}$, CEP=0) and for the reconstructed (blue open squares) fields.}
\label{Fig2}
\end{figure}

The strength of the effect and, therefore, the reconstruction of the electric field strongly depends on the relative orientation between the direction of the electronic motion (determined by the direction of polarisation of the driving field) and the instantaneous polarisation direction of the unknown field. This is demonstrated in Fig.~\ref{Fig2}a, which shows a linearly polarised field (continuous black line) and the reconstructed fields in the case of polarisation parallel (blue open square) and perpendicular (green open circle) to the motion of the electronic wave packet. The reconstruction of a (almost) zero field in the perpendicular case reveals that the electronic motion is (almost) unaffected by the unknown field in this configuration. This can be explained by observing that the phase~$S$ accumulated by the electronic wave packet depends on the integral of the kinetic energy between the ionisation and recombination instants~\cite{PRA-Lewenstein-1994}:
\begin{equation}
S\propto\int_{t_{ion}}^{t_{rec}}[\mathbf{p_s}+e\mathbf{A}_{dr}(t)+e\mathbf{A}_{unk}(t)]^2 dt\simeq\int_{t_{ion}}^{t_{rec}}[\mathbf{p_s}+e\mathbf{A}_{dr}(t)]^2+2[\mathbf{p_s}+e\mathbf{A}_{dr}(t)]\cdot e\mathbf{A}_{unk}(t)dt
\label{Eq_phase}
\end{equation}
where  $e$, $\mathbf{p_s}$, $\mathbf{A}_{dr}(t)$, and $\mathbf{A}_{unk}(t)$ indicate the electron charge, the stationary momentum, and the vector potential of the driving and unknown fields, respectively. The driving field steers the motion of the electronic wave packet along its polarisation direction ($\mathbf{p_s}$ is parallel to $\mathbf{A}_{dr}(t)$ with excellent approximation), thus making negligible the contribution of the second term of Eq.~\ref{Eq_phase} and the influence of the unknown, perturbing field, when the two pulses are perpendicularly polarised. This conclusion holds true up to intensities for which the unknown field cannot be considered a perturbation of the driving one, any longer.
As the electronic motion is sensitive to the parallel, but not to the perpendicular perturbation, the phase and amplitude of the isolated attosecond pulse encode information only on the projection of the instantaneous electric field along the direction of motion. For the reconstruction of the unknown field, the electronic dynamics acts as a time-gated directional field detector, which is sensitive only to the field component along its main axis (which coincides with the direction of motion of the electron) within a time-gating window.
Due to the dynamics of the electronic wave packet in the continuum, the duration of the gate amounts to a few hundreds of attoseconds.\\
This conclusion is confirmed in Fig.~\ref{Fig2}b, which shows the maximum of the projection of the linearly polarised field of Fig.~\ref{Fig2}a, along a reference direction ($x$), as a function of the angle~$\eta$ between this direction and the polarisation axis of the field.
As expected, the curve evolves as~$|\cos(\eta)|$ (orange line), with a small difference between $\eta=0^{\circ}$ and~$\eta=180^{\circ}$ due to the asymmetry of the few-cycle field.\\
The maximum field of the pulse, reconstructed by the analysis of the evolution of the interference pattern (magenta points), as a function of the angle~$\theta$ between the direction of the electronic wave packet and the perturbing field, matches perfectly the evolution of the maximum of the projected field. We also verified that, not only the maximum amplitudes, but also the evolution in time of the complete field is well reproduced for all relative angles.\\
The component of the electric field perpendicular to the electronic motion does neither affect the reconstruction for a generic polarisation state of the unknown pulse. This is demonstrated by the reconstruction of the field for a pulse with circular polarisation (red full circle), shown in Fig.~\ref{Fig2}a. Also in this case, the reconstructed field is given by the projection of the vectorial field along the axis of the time-gated directional field detector. A similar conclusion holds true also for elliptically polarised pulses.\\
These observations open the possibility for the characterisation of pulses with a complex time-dependent polarisation, by applying the reconstruction along two (perpendicular) polarisation directions of the isolated attosecond pulses. As an example (see Fig.~\ref{Fig2}c), we considered two perpendicularly polarised few-cycle pulses with different carrier frequencies. Due to the different periods of oscillation, the polarisation state of the total field varies from circular on the edges of the pulses (with opposite helicities) to linear in the centre. The reconstruction of the pulse along perpendicular directions ($x$ and $y$; blue squares) follows perfectly the input fields, thus giving access to the complete characterisation of the vectorial field.
\subsection{Experiment and full reconstruction of linearly polarised electric fields}
\begin{figure}[htb]
\centering\includegraphics[width=16cm]{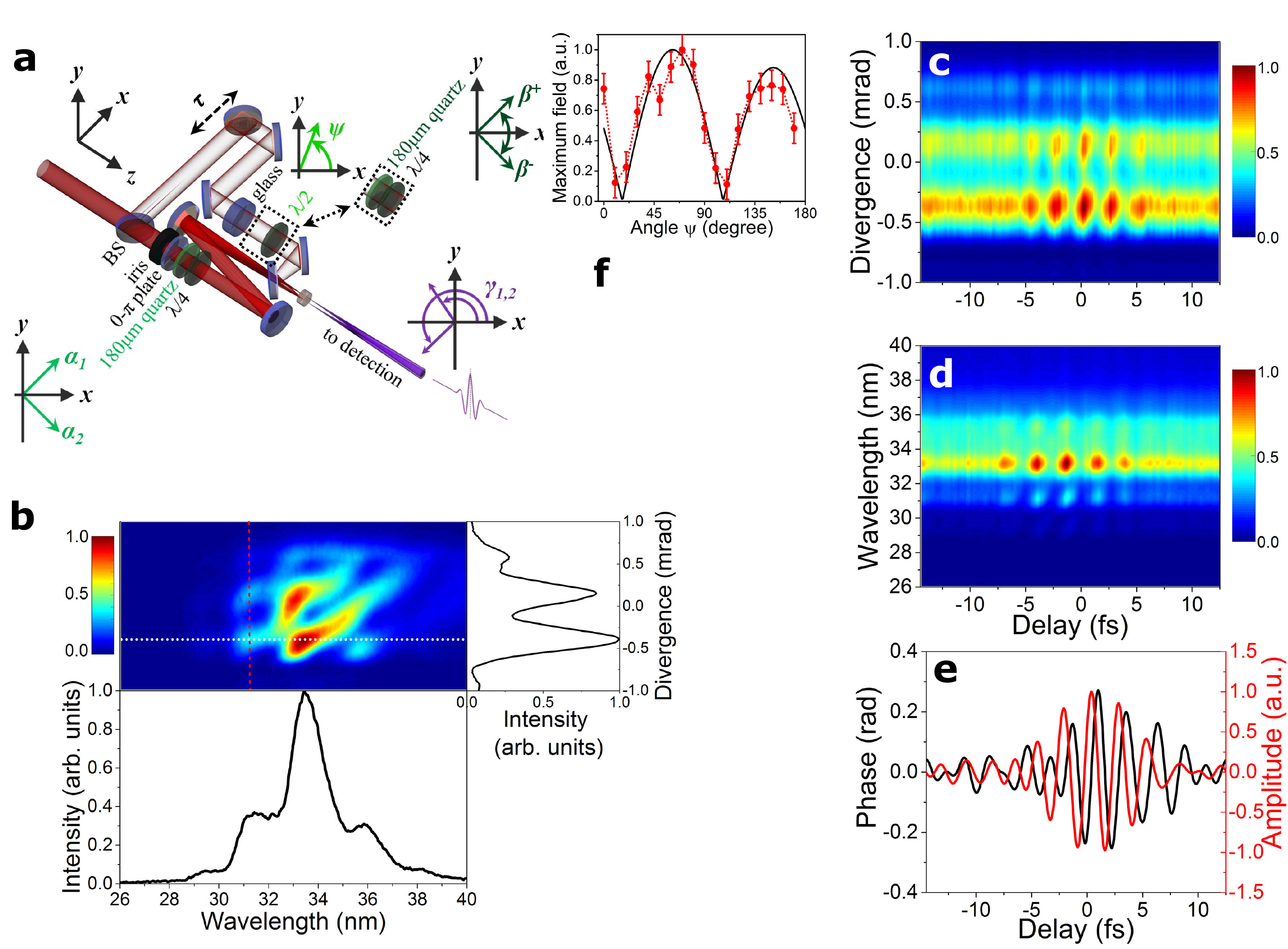}
\caption{\textbf{Experimental setup and reconstruction of linearly polarised few-cycle pulses.}\\
\textbf{a}, Experimental setup. A beam-splitter (BS) divided the incoming beam in the driving and the unknown, perturbing pulses. The intensity of the field driving the high-order harmonic generation process was adjusted by an iris. The binary 0-$\pi$ plate introduced a $\pi$-phase shift leading to two coherent foci in the focal plane of the 150-mm-focal-length focusing mirror. The polarisation of the driving pulse was modulated in time by a PG unit (delay plate with a thickness $d=180\mu m$ and zero-order quarter-wave plate). The unknown field is collinearly recombined using a drilled mirror. \textbf{b}, Experimental XUV spectral interference pattern measured in the far-field. The right and lower panel show the projections along the vertical (red dashed line) and horizontal (white dashed line) directions, corresponding to the divergence and the spectrum of the XUV interference pattern, respectively. \textbf{c},\textbf{d}, Dependence of the divergence (\textbf{c}) and of the spectrum (\textbf{d}) of the experimental XUV interference pattern as a function of the delay $\tau$. \textbf{e}, Electric fields retrieved from the amplitude (red line) and the phase (black line) using a Fourier analysis of the oscillations of \textbf{c}. \textbf{f}, Evolution of the experimental maximum electric field as function of the rotation angle $\psi$ of the $\lambda/2$ wave plate on the unknown field (red circles). The black line indicates the expected maximum of the electric field. The error bars were estimated as the amplitude of the noise on the reconstruction for delays much larger than the pulse duration.}
\label{Fig3}
\end{figure}

In the experimental setup (see Fig.~\ref{Fig3}a), we used the polarisation gating (PG) technique to implement XUV spectral interferometry with isolated attosecond pulses~\cite{NATPHYS-Sola-2006, Science-Sansone-2006}.
The few-cycle pulses passed through a binary plate, which imposed a~$\pi$ phase difference between the lower and upper halves of the beam~\cite{PRA-Camper-2014}. In the focus, this phase difference caused two closely-spaced coherent focal spots, which were the source points of the two isolated attosecond pulses.
The far-field interference pattern of the two XUV pulses is presented in Fig.~\ref{Fig3}b.
Clear interference fringes were observed in the vertical direction, while the spectrum was dispersed in the horizontal direction by a grazing incidence XUV spectrometer. The deviation of the interference fringes from the horizontal direction can be attributed to small misalignments in the grazing incidence spectrometer.\\
The unknown, perturbing field $\mathbf{E}_{unk}$ (initially linearly polarised along the $x$ direction) was overlapped with one of the two foci.
In the experiment, we typically used unknown pulses with energies on the order of a few hundreds of nanojoule corresponding to intensities of about $10^{10}-10^{11}~\mathrm{W/cm^2}$. A peak intensity as low as a few $\sim10^9~\mathrm{W/cm^2}$ still produced a significant shift of the fringes ($\sim0.1~\mathrm{rad}$), which could be experimentally measured with our setup.\\
Figures~\ref{Fig3}c and \ref{Fig3}d show the divergence and the XUV signal as a function of the relative delay between the driving and perturbing fields, respectively. The delay-dependence are in good agreement with the simulations presented in Figs.~\ref{Fig1}c,d (see Movie~2 in Supplementary Information). The field retrieved from the amplitude and phase modulations is shown in Fig.~\ref{Fig3}e. These measurements demonstrate the reconstruction of few-cycle linearly polarised pulses using spectral interferometry based on isolated attosecond pulses. The robustness of the reconstruction was verified using different XUV wavelengths and different algorithms (see Supplementary Information).\\
To demonstrate that the dynamics of the electronic wave packet is sensitive only to the weak field parallel to its motion, we measured the unknown field for different rotation angle~$\psi$ of a half-wave plate inserted in the beam path (see Fig.~\ref{Fig3}a). The maxima of the reconstructed field (red circles in Fig.~\ref{Fig3}f) well reproduced, the expected evolution~($|\cos(2\psi+\psi_0)|$, where $\psi_0$ is an offset angle), confirming that the isolated attosecond pulse encodes information only on the perturbing field parallel to its polarisation.
The lowest intensity of the component of the perturbing field along the electronic direction used in Fig.~\ref{Fig3}f was estimated in $I_{unk}=~1.5\times10^{9}~\mathrm{W/cm^2}$.
\subsection{Full reconstruction of electric fields with time-dependent polarisation}
The reconstruction of complex vectorial fields requires the characterisation of two (usually perpendicular) components. The PG technique offers a straightforward way to implement the reconstruction of optical fields along different directions, because the polarisation directions~$\gamma_{1,2}$ of the isolated attosecond pulses can be modified by simply rotating the angle of the first plate of the PG unit from~$\alpha_1=+45^{\circ}$ to~$\alpha_2=-45^{\circ}$ (see Fig.~\ref{Fig3}a and Supplementary Information).
This approach presents the advantage that the axis of our directional field sensitive detector can be rotated without any additional dispersive optical element neither on the driving nor on the perturbing field. By measuring the projections~$E_1(t)$ and~$E_2(t)$ of the vector field $\mathbf{E}_{unk}(t)$ along the two directions identified by $\gamma_1$ and $\gamma_2$, the two perpendicular components~$E_x(t)$ and~$E_y(t)$ can be reconstructed (see Supplementary Information).\\
To demonstrate the full capability of the technique, we synthesised fields with a complex time-dependent polarisation state, by replacing the half-wave plate with a replica of the PG unit on the unknown pulse (see Fig.~\ref{Fig3}a). The axis of the plates of this second PG unit were initially aligned along the polarisation of the incoming linearly polarised field. By rotating the angle~$\beta$ of the first plate to $\beta^{\pm}=\pm 45^{\circ}$, a sharp transition from circular (elliptical) to linear and back to circular(elliptical) is expected. For opposite values of~$\beta$, the helicity of the transition is expected to reverse~\cite{PRA-Sansone-2009b, NJP-Sansone-2008}. For each $\beta^{\pm}$ angle, two measurements for $\alpha_{1}$ and $\alpha_2$ were acquired.
\begin{figure}[htb]
\centering\includegraphics[width=16cm]{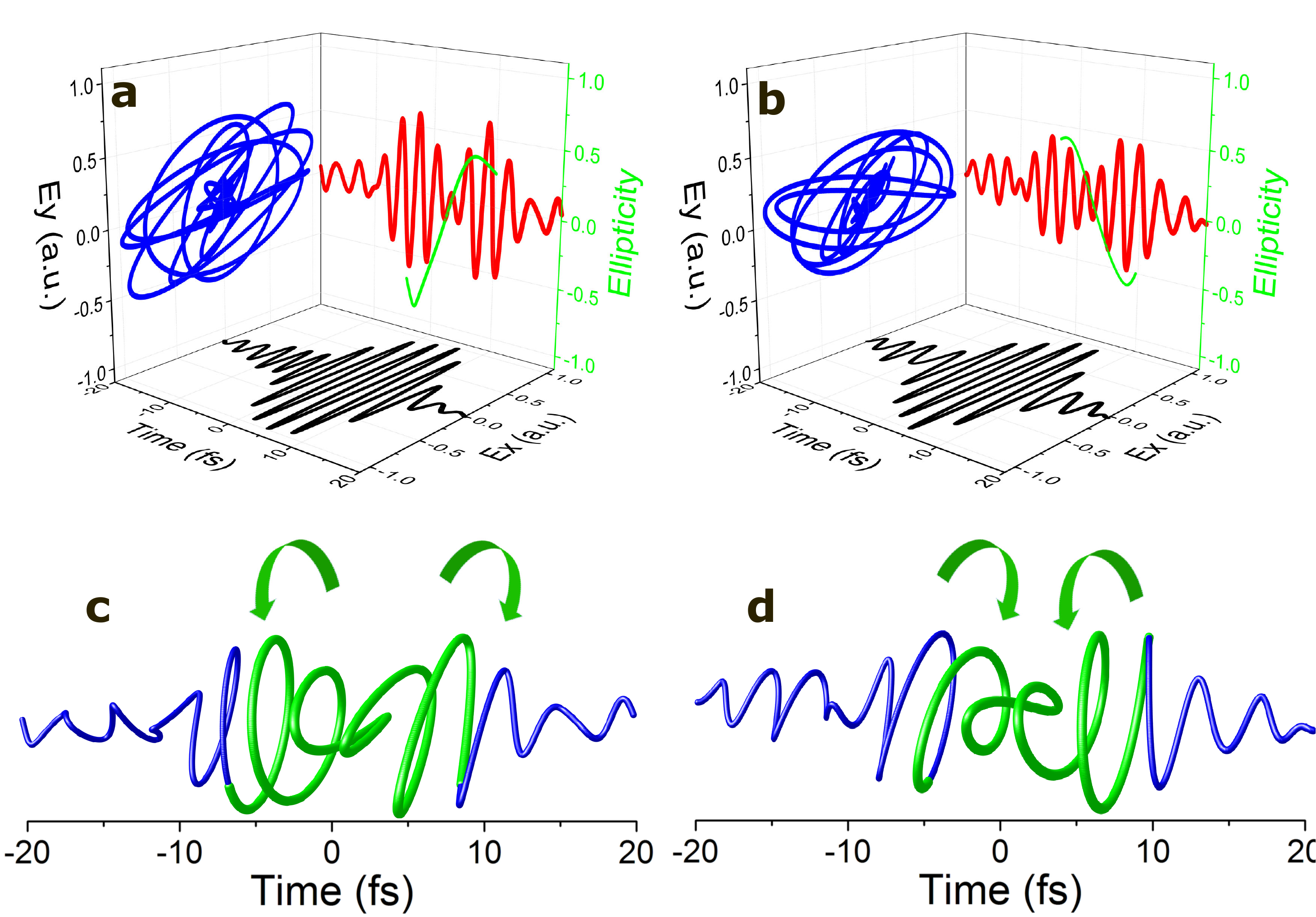}
\caption{\textbf{Reconstruction of electric fields with time-dependent polarisation.}\\
\textbf{a},\textbf{b}, Reconstructed electric fields ($E_x$ component black solid line; $E_y$ component red solid line), projection in the plane $xy$ perpendicular to the propagation direction (blue solid line), and ellipticity (green solid line) for $\beta^{+}=+45{^\circ}$ (\textbf{a}) and $\beta^{-}=-45^{\circ}$ (\textbf{b}). \textbf{c}, \textbf{d}, Three dimensional representation of the two fields with time-dependent polarisation for $\beta^{+}=+45{^\circ}$ (\textbf{c}) and $\beta^{-}=-45^{\circ}$ (\textbf{d}). The arrows indicate schematically the helicities for the two cases. The green sections correspond to the ellipticity variations shown in \textbf{a},\textbf{b}.}
\label{Fig4}
\end{figure}
Figures~\ref{Fig4}a,b show the $x$ and $y$ components of vectorial electric field retrieved for $\beta^{+}$ and $\beta^{-}$, respectively, and the projection of the field in the $x-y$ plane.
The pulse was characterised by a complex time-dependent ellipticity (see Figs.~\ref{Fig4}a,b), which evolves from elliptical (positive helicity) to linear and back to elliptical (negative helicity) for $\beta^{+}=+45^{\circ}$.
The transition between different polarisation states can be clearly observed in the three-dimensional representation of the fields, which evidences a rotation of the polarisation vector clockwise (counter clockwise) on the leading (falling) edge of the pulse for $\beta^{+}$ (Fig.~\ref{Fig4}c)~\cite{PRA-Antoine-1996}. For $\beta^{-}$, the rotation directions are reversed (Fig.~\ref{Fig4}d), as expected.\\
Our approach also allows for the characterisation of the carrier envelope phase (CEP) of the field, by analysing the variations of the electric field components in the time window of linear polarisation. The CEP was controlled by changing the amount of fused silica inserted in the beam path~\cite{PRL-Paulus-2003}. The reconstruction for three CEPs spaced by $\pi/2$ is shown in Figs.~\ref{Fig5}a,b,c. We can observe that the ellipticity is only slightly affected by the CEP, as well as the component of the electric field along the $x$ direction (apart from a shift in time). The minor component of the field, on the other hand, is strongly modified and it is even reversed for a $\pi$ phase shift (Figs.~\ref{Fig5}a,c). The simulations are shown in Figs.~\ref{Fig5}d,e,f, which report the field for the three CEPs $\varphi=0$ (d), $\varphi=\pi/2$ (e), and $\varphi=\pi$ (f).
The good matching between the simulations and the experiments allows one for a reliable determination of the CEP of the unknown field.
The measured projection in the $xy$ plane is also reported in Figs.~\ref{Fig5}g,h,i. The field corresponding to the time window of Figs.~\ref{Fig5}a,b,c  is shown in green. The transition through a state of linear polarisation is clearly visible, and it evidences that the projections are characterised by narrow (and opposite) region of linear polarisation for $\varphi=0,\pi$. This transition region becomes broader for intermediate CEPs ($\varphi=\pi/2$). This periodic variation is in excellent agreement with simulations (see Movie~3 in Supplementary Information).\\
\begin{figure}[htb]
\centering\includegraphics[width=16cm]{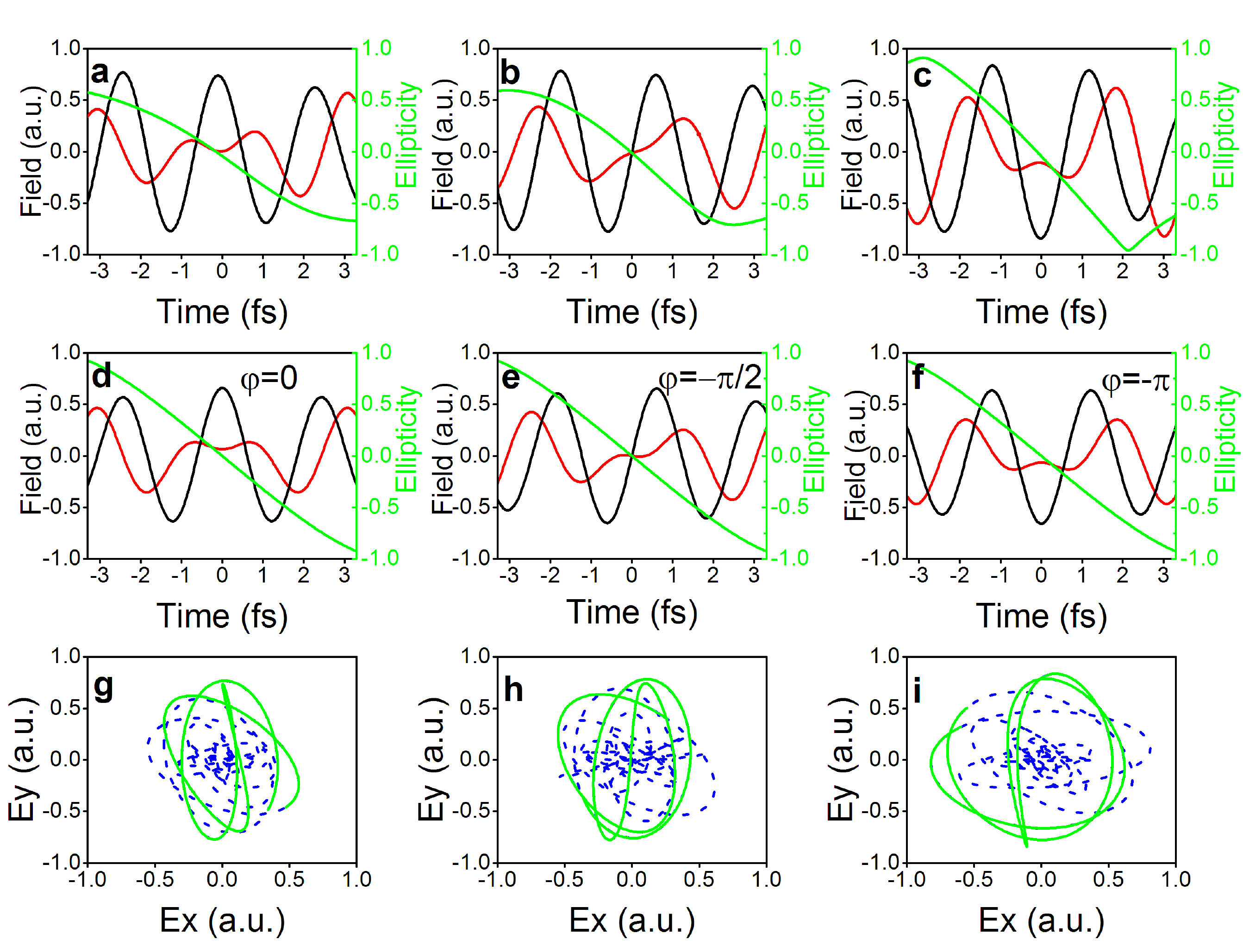}
\caption{\textbf{Effect of the CEP in the temporal window of linear polarisation.}
Measured (\textbf{a},\textbf{b},\textbf{c}) and simulated (\textbf{d},\textbf{e},\textbf{f}) electric fields ($E_x$ component black solid line; $E_y$ component red solid line) and ellipticities (green solid line) for $\beta^{+}=+45^{\circ}$ and three CEPs $\varphi=0$ (\textbf{a},\textbf{d}), $\varphi=\pi/2$ (\textbf{b},\textbf{e}), and $\varphi=\pi$ (\textbf{c},\textbf{f}). The ellipticity in the central temporal window (-3.3~fs-3.3~fs) is highlighted in green.
Experimental projection (blue dashed line) of the electric fields perpendicular to the propagation direction for the three CEPs $\varphi=0$ (\textbf{g}), $\varphi=\pi/2$ (\textbf{h}), and $\varphi=\pi$ (\textbf{i}). The green lines correspond to the projections in the temporal window (-3.3~fs-3.3~fs), shown in \textbf{a},\textbf{b},\textbf{c}.}
\label{Fig5}
\end{figure}
\subsection{Conclusions}
In conclusion, we have demonstrated the complete temporal characterisation of the time-dependent amplitudes of the two components (including their carrier-envelope phases) of a few-cycle pulse with a complex polarisation state. The experimental approach is based on the implementation of spectral XUV interferometry and on the demonstration that the dynamics of a single electron wave packet acts as an attosecond directional sensitive field detector. These results opens the possibility for the complete characterisation of an optical field with a generic polarisation state using an all-optical approach free from distortions. Combined with phase sensitive detection techniques based on lock-in amplifier, the sensitivity of the techniques can be further improved opening new perspectives for the full characterisation (amplitude, phase, and carrier-envelope phase) of probe fields used in pump-probe spectroscopy. This approach would offer the possibility to reconstruct the complete response (complex dielectric wave function) of metallic and nanostructured system in the visible and infrared spectral range, in perfect analogy to time-domain spectroscopy for semiconductors in the THz domain.
\subsection{Methods}
The data that support the plots within this paper and other findings of this study are available from the corresponding author upon reasonable request.

\subsection{Acknowledgements}
This project has received also funding from the European Union's Horizon 2020 research and innovation programme under the Marie Sklodowska-Curie grant agreement no.~641789 Molecular Electron Dynamics investigated by Intense Fields and Attosecond Pulses (MEDEA), the European Research Council Starting Grant agreement no.~637756 Steering attosecond electron dynamics in biomolecules with UV-XUV LIGHT pulses (STARLIGHT). G.~G.~P. acknowledges support from the German Science Foundation (PA 730/7 within priority programme 1840).
\subsection{Contributions}
P.~C., M.~R., and G.~S. conceived and planned the experiment. P.~C., M.~R., and H.~A. conducted the experiment. M.~R. and A.~C. analysed the data. S.~K., F.~C, and M.~N., contributed to the development of the experimental setup. L.~P. and F.~F. designed and installed the XUV beamline. D.~H and G.~G.~P. designed and installed the STEREO ATI. J.~U., C.~D.~S and R.~M. designed and installed the Reaction Microscope. G.~S. performed the simulations and theoretical analysis. The manuscript was drafted by G.~S. and completed in consultation with all authors.

\subsection{Competing financial interests}
The authors declare no competing financial interests.

Correspondence and requests for materials should be addressed to G.~Sansone\\
(giuseppe.sansone@physik.uni-freiburg.de\\giuseppe.sansone@polimi.it).

\end{document}